# Effect of laser on thermopower of chiral carbon nanotube

S.Y.Mensah[a], A. Twum[a], N. G. Mensah[b], K. A. Dompreh[a], S. S. Abukari[a], and G. Nkrumah-Buandoh[c]

[a]Department of Physics, Laser and Fibre Optics Centre, University of Cape Coast, Cape Coast, Ghana

[b]Department of Mathematics, University of Cape Coast, Cape Coast, Ghana

[c]University of Ghana, Legon, Ghana

*Corresponding author. S. Y. Mensah

Tel.:+233 042 33837

E-mail address: profsymensah@yahoo.co.uk

**Abstract**

An investigation of laser stimulated thermopower in chiral CNT is presented. The thermopower of a chiral CNT is calculated using a tractable analytical approach. This is done by solving the Boltzmann kinetic equation with energy dispersion relation obtained in the tight binding approximation to determine the electrical and thermal properties of chiral carbon nanotubes. The differential thermoelectric power along the circumferential $\alpha_{cz}$ and axial $\alpha_{zz}$ are obtained. The results obtained are numerically analyzed and $\alpha$ is found to oscillate in the presence of laser radiations. We have also noted that Laser source above $4.6 \times 10^7$ V/m lowered the thermopower otherwise there is no change. Varying $\Delta_s$ and $\Delta_z$ the thermopower changes from positive to negative.

1. **Introduction**

It is well known that a circuit made from two dissimilar metals, with junctions at different temperatures induces an electrical current in the circuit. The temperature difference produces an electric potential (voltage) which can drive an electric current in a closed circuit. The voltage produced is proportional to the temperature difference between the two junctions. The proportionality constant $S$ is defined as the Seebeck coefficient or thermoelectric power and is obtained from the ratio of the voltage generated $\Delta V$ to the applied temperature difference $\Delta T$ (i.e. $S = \Delta V / \Delta T$) [1]. Lyeo *et al.* in [2] have reported on an experimental technique called Scanning Thermoelectric Microscopy (SThEM) that can probe thermoelectric transport at nanoscales. Reference [2] demonstrated this by mapping out the thermopower of a *pn* homojunction and had a remarkable result which showed the spatial resolution to be on the order of 2 to 4 nm in highly doped semiconductors. This creates the possibility of probing semiconductor nanostructures for thermoelectricity. Interestingly, this resolution is of the order of the nanostructure size discovered by Hsu *et al.* [3] for $AgPb_{18}SbTe_{20}$.

Thermoelectric (TE) power has been reported for a random array of carbon nanotubes (CNT) [4-6] as well as for individual tubes [7]. Similar investigations were made on quantum wires [8, 9] and artificial nanostructures, such as superlattices [10]. Mensah and Buah-Bassuah [11] have investigated the photostimulated thermomagnetic effect by electrons in a semiconductor superlattice (SL). They indicated the possibility of controlling the thermopower α, the electron thermal conductivity χ, and the electroconductivity σ of the SL with the help of laser radiation. They found the parameters α, χ, and σ to oscillate in the presence of laser therefore are amplitude dependent.

Past work on CNT was mostly made on randomly dispersed tubes but Shamim M. et al. [12] reported on the TE properties of cross aligned and co aligned junctions made between functionalized single-wall CNTs (SWCNTs) and multiwall CNTs (MWCNTs).

Mensah *et al*. have studied the differential thermopower of the chiral carbon nanotube [13]. They used the approach stated in [14] together with the model developed in [15] to determine the thermopower $\alpha$ of the chiral CNT. The approach requires the creation of phenomenological models that yield analytically tractable results [15]. The justification for this approach can be established from the work of Miyamoto *et al*. [16], where they computed the current excited in carbon and $BC_2N$ nanotubes immersed in an electrostatic field. They observed in [13] that the thermopower strongly depends on the geometric chiral angle (GCA) $\theta_h$, electric field E, temperature *T,* the real overlapping integrals for jumps along the tubular axis $\Delta_z$ and the base helix $\Delta_s$ . In this work, we will use the approach in [13] to investigate theoretically the laser stimulated thermopower in chiral CNTs.

The paper is organized as follows: section one deal with the introduction; in section two, we establish the theory and solutions; results obtained will be discussed in section three and finally we draw our conclusions.

## 2. Theory

The carrier (electron or hole) current density j, electrical conductivity σ and thermopower α of a chiral SWNT are calculated as functions of the geometric chiral angle $\theta_h$, temperature T, the real overlapping integrals for jumps along the nanotube axis $\Delta_z$ and along the base helix $\Delta_s$. The calculation is done using the approach in reference [14] together with the phenomenological model of a SWNT developed in references [15] and [17]. This model yields physically interpretable results and gives correct qualitative descriptions of various electronic processes, which are corroborated by the first-principle numerical simulations of Miyamoto et al [16].

Following the approach of [13], we consider a SWNT under a temperature gradient $\nabla T$ and in an electric field applied along the nanotube axis. Employing the Boltzmann kinetic equation

$$\frac{\partial f(r,p,t)}{\partial t} + v(p)\frac{\partial f(r,p,t)}{\partial r} + eE(t)\frac{\partial f(r,p,t)}{\partial p} = \frac{\partial (r,p,t) - f_0(p)}{\tau} \qquad (1)$$

where *f(r, p, t)* is the distribution function, *f₀(p)* is the equilibrium distribution function, *v(p)* is the electron velocity, $E(t) = E_0 + E_1 \cos(wt)$ is the magnitude of the electric field, with $E_0$ being constant electric field and $E_1 \cos(wt)$ being monochromatic laser source, *r* is the electron position, *p* is the electron dynamical momentum, *t* is time elapsed, *τ* is the electron relaxation time and *e* is the electron charge and taken the collision integral in the *τ* approximation and further assumed constant, the exact solution of Equation (1) is solved using perturbation approach where the second term is treated as the perturbation. In the linear approximation of $\nabla T$ and $\nabla \mu$, the solution to the Boltzmann kinetic equation is

$$f(p,t) = \tau^{-1} \int_0^\infty \exp\left(-\frac{t}{\tau}\right) f_0 \left( p - e \int_{t-t'}^{t} [E_0 + E_1 \cos wt''] dt'' \right) dt$$

$$+ \int_0^\infty \exp\left(-\frac{t}{\tau}\right) dt \left\{ \left[ \varepsilon \left( p - e \int_{t-t'}^{t} [E_0 + E_1 \cos wt''] dt'' \right) - \mu \right] \frac{\nabla T}{T} + \nabla \mu \right\}$$

$$\times v \left( p - e \int_{t-t'}^{t} [E_0 + E_1 \cos wt''] dt'' \right) \frac{\partial f_0}{\partial \varepsilon} \left( p - e \int_{t-t'}^{t} [E_0 + E_1 \cos wt''] dt'' \right) \qquad (2)$$

$\varepsilon(p)$ is the tight-binding energy of the electron, and $\mu$ is the chemical potential. The carrier current density j is defined as

$$j(t) = e \sum_p v(p) f(p,t) \qquad (3)$$

Substituting Eq. (2) into Eq. (3) we have

$$j(t) = e\tau^{-1} \int_0^\infty \exp\left(-\frac{t}{\tau}\right) dt \sum_p v(p) f_0\left(p - e\int_{t-t'}^t [E_0 + E_1 \cos wt''] dt''\right)$$

$$+ e \int_0^\infty \exp\left(-\frac{t}{\tau}\right) dt \sum_p v(p) \left\{ \left[\varepsilon\left(p - e\int_{t-t'}^t [E_0 + E_1 \cos wt''] dt''\right) - \mu\right] \frac{\nabla T}{T} + \nabla \mu \right\}$$

$$\times v\left(p - e\int_{t-t'}^t [E_0 + E_1 \cos wt''] dt''\right) \frac{\partial f_0}{\partial \varepsilon}\left(p - e\int_{t-t'}^t [E_0 + E_1 \cos wt''] dt''\right) \qquad (4)$$

Making the transformation

$$p - e\int_{t-t'}^t [E_0 + E_1 \cos wt''] dt'' \to p,$$

we obtain for the current density

$$j(t) = e\tau^{-1} \int_0^\infty \exp\left(-\frac{t}{\tau}\right) dt \sum_p v\left(p - e\int_{t-t'}^t [E_0 + E_1 \cos wt''] dt''\right) f_0(p)$$

$$+ e \int_0^\infty \exp\left(-\frac{t}{\tau}\right) dt \sum_p \left\{ [\varepsilon(p) - \mu] \frac{\nabla T}{T} + \nabla \mu \right\}$$

$$\times \left\{ v(p) \frac{\partial f_0(p)}{\partial \varepsilon} \right\} v\left(p - e\int_{t-t'}^t [E_0 + E_1 \cos wt''] dt''\right) \qquad (5)$$

Using the phenomenological model [15,17,18], a SWNT is considered as an infinitely long periodic chain of carbon atoms wrapped along a base helix and the current density is written in the form

$$j = S' u_s + Z' u_z \qquad (6)$$

where S′ and Z′ are respectively components of the current density along the base helix and along the nanotube axis. The motion of electrons in the SWNT is resolved along the nanotube axis in the direction of the unit vector $u_z$ and a unit vector $u_s$ tangential to the base helix. $u_c$ is defined as the unit vector tangential to the circumference of the nanotube and $\theta_h$ is the geometric chiral angle (GCA). $u_c$ is always perpendicular to $u_z$, therefore $u_s$ can be resolved along $u_c$ and $u_z$ as

$$u_s = u_c \cos\theta_h + u_z \sin\theta_h \qquad (7)$$

Therefore, j can be expressed in terms of $u_c$ and $u_z$ as

$$j = u_c (S' \cos\theta_h) + u_z (Z' + S' \sin\theta_h) \equiv j_c u_c + j_z u_z \qquad (8)$$

which implies that,

$$j_c = S' \cos\theta_h \qquad (9)$$

$$j_z = Z' + S' \sin \theta_h \tag{10}$$

The interference between the axial and helical paths connecting a pair of atoms is neglected so that transverse motion quantization is ignored [15,17]. This approximation best describes doped chiral carbon nanotubes, and is experimentally confirmed in [19].
Thus if in Eq (5) the transformation

$$\sum_p \to \frac{2}{(2\pi\hbar)^2} \int_{\pi/d_s}^{\pi/d_s} dP_s \int_{\pi/d_z}^{\pi/d_z} dP_z$$

is made, $Z'$ and $S'$ respectively become,

$$Z' = \frac{2e\tau^{-1}}{(2\pi\hbar)^2} \int_0^\infty \exp\left(-\frac{t}{\tau}\right) dt \int_{\pi/d_s}^{\pi/d_s} dP_s \int_{\pi/d_z}^{\pi/d_z} dP_z v_z\left(p - e\int_{t-t'}^t [E_0 + E_1 \cos wt''] dt''\right) f_0(p)$$

$$+ \frac{2e}{(2\pi\hbar)^2} \int_0^\infty \exp\left(-\frac{t}{\tau}\right) dt \int_{\pi/d_s}^{\pi/d_s} dP_s \int_{\pi/d_z}^{\pi/d_z} dP_z \left\{ [\varepsilon(p) - \mu] \frac{\nabla_z T}{T} + \nabla_z \mu \right\}$$

$$\times \left\{ v_z(p) \frac{\partial f_0(p)}{\partial \varepsilon} \right\} v_z\left(p - e\int_{t-t'}^t [E_0 + E_1 \cos wt''] dt''\right) \tag{11}$$

and

$$S' = \frac{2e\tau^{-1}}{(2\pi\hbar)^2} \int_0^\infty \exp\left(-\frac{t}{\tau}\right) dt \int_{\pi/d_s}^{\pi/d_s} dP_s \int_{\pi/d_z}^{\pi/d_z} dP_z v_s\left(p - e\int_{t-t'}^t [E_0 + E_1 \cos wt''] dt''\right) f_0(p)$$

$$+ \frac{2e}{(2\pi\hbar)^2} \int_0^\infty \exp\left(-\frac{t}{\tau}\right) dt \int_{\pi/d_s}^{\pi/d_s} dP_s \int_{\pi/d_z}^{\pi/d_z} dP_z \left\{ [\varepsilon(p) - \mu] \frac{\nabla_s T}{T} + \nabla_s \mu \right\}$$

$$\times \left\{ v_s(p) \frac{\partial f_0(p)}{\partial \varepsilon} \right\} v_s\left(p - e\int_{t-t'}^t [E_0 + E_1 \cos wt''] dt''\right) \tag{12}$$

where the integrations are carried out over the first Brillouin zone, $\hbar$ is Planck's constant, $v_s$, $p_s$, $E_s$, $\nabla_s T$, and $\nabla_s \mu$ are the respective components of $v$, $p$, $E$, $\nabla T$ and $\nabla \mu$ along the base helix, and $v_z$, $p_z$, $E_z$, $\nabla_z T$, and $\nabla_z \mu$ are the respective components of $v$, $p$, $E$, $\nabla T$ and $\nabla \mu$ along the nanotube axis.
The energy dispersion relation for a chiral nanotube obtained in the tight binding approximation [16] is

$$\varepsilon(p) = \varepsilon_0 - \Delta_s \cos \frac{P_s d_s}{\hbar} - \Delta_z \cos \frac{P_z d_z}{\hbar} \tag{13}$$

where $\varepsilon_o$ is the energy of an outer-shell electron in an isolated carbon atom, $\Delta_z$ and $\Delta_s$ are the real overlapping integrals for jumps along the respective coordinates, $p_s$ and $p_z$ are the components of momentum tangential to the base helix and along the the nanotube axis, respectively. The components $v_s$ and $v_z$ of the electron velocity $v$ are respectively calculated from the energy dispersion relation Eq (13) as

$$v_s(p) = \frac{\partial \varepsilon(p)}{\partial P_s} = \frac{\Delta_s d_s}{\hbar} \sin \frac{P_s d_s}{\hbar} \tag{14}$$

$$v_s\left(p - e\int_{t-t'}^{t}[E_0 + E_1 \cos wt''] dt''\right) = \frac{\Delta_s d_s}{\hbar} \sin\left(p - e\int_{t-t'}^{t}[E_0 + E_1 \cos wt''] dt''\right)$$

$$= \frac{\Delta_s d_s}{\hbar}\left\{\sin\frac{P_s d_s}{\hbar} \cos\left(p - e\int_{t-t'}^{t}[E_0 + E_1 \cos wt''] dt''\right)\right.$$

$$\left. - \cos\frac{P_s d_s}{\hbar} \sin\left(p - e\int_{t-t'}^{t}[E_0 + E_1 \cos wt''] dt''\right)\right\} \tag{15}$$

$$v_z(p) = \frac{\partial \varepsilon(p)}{\partial P_z} = \frac{\Delta_z d_z}{\hbar} \sin \frac{P_z d_z}{\hbar} \tag{16}$$

and

$$v_z\left(p - e\int_{t-t'}^{t}[E_0 + E_1 \cos wt''] dt''\right) = \frac{\Delta_z d_z}{\hbar}\left\{\sin\frac{P_s d_s}{\hbar} \cos\left(p - e\int_{t-t'}^{t}[E_0 + E_1 \cos wt''] dt''\right)\right.$$

$$\left. - \cos\frac{P_z d_z}{\hbar} \sin\left(p - e\int_{t-t'}^{t}[E_0 + E_1 \cos wt''] dt''\right)\right\} \tag{17}$$

To calculate the carrier current density for a non-degenerate electron gas, the Boltzmann equilibrium distribution function $f_0(p)$ is expressed as

$$f_0(p) = C \exp\left(\frac{\Delta_s \cos\frac{P_s d_s}{\hbar} + \Delta_z \cos\frac{P_z d_z}{\hbar} + \mu - \varepsilon_0}{kT}\right) \tag{18}$$

Where C is found to be

$$C = \frac{d_s d_z n_0}{2 I_0(\Delta_s^*) I_0(\Delta_z^*)} \exp\left(-\frac{\mu - \varepsilon_0}{kT}\right) \quad (19)$$

and $n_0$ is the surface charge density, $I_n(x)$ is the modified Bessel function of order n defined by

$\Delta_s^* = \frac{\Delta_s}{kT}$ and $\Delta_z^* = \frac{\Delta_z}{kT}$ and $k$ is Boltzmann's constant.

Now, we substituted Eqs (13) - (18) into Eqs (11) and (12), and carried out the integrals and also averages over t to obtain the following expressions

$$S' = -\sigma_s(E) E_{sn}^* - \sigma_s(E) \frac{k}{e} \left\{ \left(\frac{\varepsilon_0 - \mu}{kT}\right) - \Delta_s^* \frac{I_0(\Delta_s^*)}{I_1(\Delta_s^*)} + 2 - \Delta_z^* \frac{I_1(\Delta_z^*)}{I_0(\Delta_z^*)} \right\} \nabla_s T \quad (20)$$

$$Z' = -\sigma_z(E) E_{zn}^* - \sigma_z(E) \frac{k}{e} \left\{ \left(\frac{\varepsilon_0 - \mu}{kT}\right) - \Delta_z^* \frac{I_0(\Delta_z^*)}{I_1(\Delta_z^*)} + 2 - \Delta_s^* \frac{I_1(\Delta_s^*)}{I_0(\Delta_s^*)} \right\} \nabla_z T \quad (21)$$

Where we have defined $E_{sn}^*$ as

$$E_{sn}^* = E_n + \nabla_s \frac{\mu}{e}$$

and also $\sigma_i(E)$ as

$$\sigma_i(E) = \frac{e^2 \tau \Delta_i d_i^2 n_0}{\hbar^2} \frac{I_1(\Delta_i^*)}{I_0(\Delta_i^*)} \sum_{n=-\infty}^{\infty} J_n^2(a) \left[ \frac{1}{1 + \left(\frac{ed_i E_0}{\hbar} + nw\right)^2 \tau^2} \right], \quad i = s, z \quad (22)$$

Here $J_n(a)$ is the Bessel function of the n order.

Substituting Eq (20) into Eq (9) gives circumferential current density $j_c$ as

$j_c = -\sigma_s(E) \sin\theta_h \cos\theta_h E_{zn}^*$

$$-\sigma_s(E) \frac{k}{e} \sin\theta_h \cos\theta_h \left\{ \left(\frac{\varepsilon_0 - \mu}{kT}\right) - \Delta_s^* \frac{I_0(\Delta_s^*)}{I_1(\Delta_s^*)} + 2 - \Delta_z^* \frac{I_1(\Delta_z^*)}{I_0(\Delta_z^*)} \right\} \nabla_z T \quad (23)$$

Similarly, the axial current density $j_z$ was obtained after substituting Eq (21) into Eq (10) as,

$$j_z = -\{\sigma_z(E) + \sigma_s(E)\sin^2\theta_h\}E_{zn}^* - \left\{\sigma_z(E)\frac{k}{e}\left[\left(\frac{\varepsilon_0-\mu}{kT}\right) - \Delta_z^*\frac{I_0(\Delta_z^*)}{I_1(\Delta_z^*)} + 2 - \Delta_s^*\frac{I_1(\Delta_s^*)}{I_0(\Delta_s^*)}\right]\right.$$

$$\left. + \sigma_s(E)\frac{k}{e}\sin^2\theta_h\left[\left(\frac{\varepsilon_0-\mu}{kT}\right) - \Delta_s^*\frac{I_0(\Delta_s^*)}{I_1(\Delta_s^*)} + 2 - \Delta_z^*\frac{I_1(\Delta_z^*)}{I_0(\Delta_z^*)}\right]\right\}\nabla_z T \quad (24)$$

we define

$$\xi = \frac{\varepsilon_0-\mu}{kT}, \quad A_i = \frac{I_1(\Delta_i^*)}{I_0(\Delta_i^*)}, \quad B_i = \frac{I_0(\Delta_i^*)}{I_1(\Delta_i^*)} - \frac{2}{\Delta_i^*}, \quad i = s, z \quad (25)$$

Then Eqs (23) and (24) respectively become

$$j_c = -\sigma_s(E)\sin\theta_h\cos\theta_h E_{zn}^* - \sigma_s(E)\frac{k}{e}\sin\theta_h\cos\theta_h\{\xi - \Delta_s^*B_s - \Delta_z^*A_z\}\nabla_z T \quad (26)$$

and

$$j_z = -\{\sigma_z(E) + \sigma_s(E)\sin^2\theta_h\}E_{zn}^*$$

$$-\left\{\sigma_z(E)\frac{k}{e}[\xi - \Delta_z^*B_z - \Delta_s^*A_s] + \sigma_s(E)\frac{k}{e}\sin^2\theta_h[\xi - \Delta_s^*B_s - \Delta_z^*A_z]\right\}\nabla_z T \quad (27)$$

Eqs (26) and (27) define the carrier current density. The circumferential $\sigma_{cz}$ and axial $\sigma_{zz}$ components of the electrical conductivity in the CNT are obtained from Eqs (26) and (27) respectively. In fact the coefficients of the electric field $-E_{zn}^*$ in these equations define $\sigma_{cz}$ and $\sigma_{zz}$ as follows,

$$\sigma_{cz} = \sigma_s(E)\sin\theta_h\cos\theta_h \quad (28)$$

$$\sigma_{zz} = \sigma_z(E) + \sigma_s(E)\sin^2\theta_h \quad (29)$$

The differential thermoelectric power is defined as the ratio $\frac{|E_{zn}^*|}{|\nabla T|}$ in an open circuit (i.e. when j = 0). Thus setting $j_c$ to zero in Eq (23), the thermoelectric power $\alpha_{cz}$ along the circumferential direction is obtained as follows

$$0 = -\sigma_s(E)\sin\theta_h\cos\theta_h E_{zn}^* - \sigma_s(E)\frac{k}{e}\sin\theta_h\cos\theta_h\{\xi - \Delta_s^*B_s - \Delta_z^*A_z\}\nabla_z T$$

$$\sigma_s(E)\sin\theta_h\cos\theta_h E_{zn}^* = -\sigma_s(E)\frac{k}{e}\sin\theta_h\cos\theta_h\{\xi - \Delta_s^*B_s - \Delta_z^*A_z\}\nabla_z T$$

$$\frac{E^*_{zn}}{\nabla_z T} = -\frac{\sigma_s(E)\frac{k}{e}\sin\theta_h \cos\theta_h \{\xi - \Delta^*_s B_s - \Delta^*_z A_z\}}{\sigma_s(E)\sin\theta_h \cos\theta_h}$$

$$\alpha_{cz} = \left|\frac{E^*_{zn}}{\nabla_z T}\right| = \frac{k}{e}\{\xi - \Delta^*_s B_s - \Delta^*_z A_z\} \qquad (30)$$

Similarly, the thermoelectric power $\alpha_{zz}$ along the axial direction is obtained from Eq (43) as follows (i.e. when $j_z = 0$)

$$\left|\frac{E^*_{zn}}{\nabla_z T}\right| = -\frac{\left\{\sigma_z(E)\frac{k}{e}[\xi - \Delta^*_z B_z - \Delta^*_s A_s]\right\}}{\sigma_z(E) + \sigma_s(E)\sin^2\theta_h} + \frac{\sigma_s(E)\frac{k}{e}\sin^2\theta_h [\xi - \Delta^*_s B_s - \Delta^*_z A_z]}{\sigma_z(E) + \sigma_s(E)\sin^2\theta_h}$$

$$\alpha_{zz} = \left|\frac{E^*_{zn}}{\nabla_z T}\right| = \frac{\sigma_z(E)}{\sigma_z(E) + \sigma_s(E)\sin^2\theta_h}\frac{k}{e}[\xi - \Delta^*_z B_z - \Delta^*_s A_s]$$

$$+ \frac{\sigma_s(E)\sin^2\theta_h}{\sigma_z(E) + \sigma_s(E)\sin^2\theta_h}\frac{k}{e}[\xi - \Delta^*_s B_s - \Delta^*_z A_z] \qquad (31)$$

In summary, the analytical expressions obtained for the carrier current density $j$ and thermopower $\alpha$ depend on the geometric chiral angle $\theta_h$, temperature T, the real overlapping integrals for jumps along the tubular axis $\Delta_z$ and the base helix $\Delta_s$.

When the Laser source is switched off i.e. $E_s = 0$ and $w = 0$ the thermopower expression in Eq (31) reduces to

$$\alpha_{zz} = \frac{\sigma_z(E)}{\sigma_z(E) + \sigma_s(E)\sin^2\theta_h}\frac{k}{e}[\xi - \Delta^*_z B_z - \Delta^*_s A_s]$$

$$+ \frac{\sigma_s(E)\sin^2\theta_h}{\sigma_z(E) + \sigma_s(E)\sin^2\theta_h}\frac{k}{e}[\xi - \Delta^*_s B_s - \Delta^*_z A_z] \qquad (32)$$

where $\sigma_i(E) = \frac{e^2 \tau \Delta_i d_i^2 n_0}{\hbar^2}\frac{I_1(\Delta^*_i)}{I_0(\Delta^*_i)}\left[\frac{1}{1+\left(ed_i E_0/\hbar\right)^2 \tau^2}\right]$, $I = s, z$ .

these expressions were obtained in [13]

**Results, Discussion and Conclusion**

Using Boltzmann kinetic equation, the expressions of the carrier current density and thermopower of chiral SWNT was obtained.

We observed from Eqs (30) and (31) that the thermoelectric power of a chiral CNT is dependent on the electric fields $E_s$ and $E_o$, temperature T, GCA $\theta_h$, and the overlapping integrals $\Delta_s$ and $\Delta_z$ for jumps along the circumferential and axial directions. . To further understand how these parameters affect the thermopower, we sketched Eq (31) using MATLAB.

Figure (1a) represents the dependence of thermopower $\alpha_{zz}$ on temperature for a fixed value of $\Delta_z$ = 0.015eV and values of $\Delta_s$ varied from 0.015eV to 0.025eV. It was observed that the thermopower decreases rapidly with increasing temperature for values of $\Delta_s$ between 0.015eV and 0.018eV. For values of $\Delta_s$ above 0.018eV, the thermopower increases rapidly to a maximum value and then start decreasing gradually with increasing temperature. At high temperatures above 500K, thermopower assumes a lower constant value for all values of $\Delta_s$. A similar behavior was observed by J. Hone et al. in [19], where they measured the thermopower of a SWNT experimentally.

The hyperbolic curves obtained in Figure (1a) are similar to the characteristic thermopower behavior expected for semiconducting CNTs [20]. We noted that when $\Delta_s$ values are equal or slightly above $\Delta_z$, i.e. $\Delta_s$ values between 0.015eV and 0.020eV, the thermopower $\alpha$ decreases with increasing temperature T, which give hyperbolic curves. These conditions make the chiral CNT behaves as a semimetal. The fact that thermopower values in Figure (1a) are positive over the entire range of temperature indicates that the contribution from positive (hole) carriers dominates the response.

The dependence of thermopower on temperature is also sketched for fixed values of $\Delta_z$ = 0.024eV, 0.027eV and 0.041eV as Figures (1b), (1c) and (1d) respectively. In all cases, $\Delta_s$ is varied from 0.015eV to 0.025eV.

In Figures (1b) and (1c), the thermopower was found to increase rapidly to a maximum value, then decreases slowly to a constant value as temperature rises. All the curves were observed to have turning points at different temperatures.

Comparing our results obtained with the experimentally measured thermopower in reference [21], it was noted that the theoretical curves agree reasonably well with the experimental values. Careful study of all the curves obtained revealed that the turning points shift toward lower temperatures for a given $\Delta_z$ and increasing $\Delta_s$, but they shift towards greater temperatures as $\Delta_z$ increases.

Interestingly, it came to light that there exists a threshold temperature for which hole conductivity switches over to electron conductivity. It means that positive thermopower of the chiral CNT becomes negative. The threshold value for the temperature shifts towards lower temperature as $\Delta_z$ is increased. This can be explained by the fact that graphite has a pair of weakly overlapping electron and hole $sp^2$ or $\pi$ bands with near mirror symmetry about the Fermi energy $E_F$. Approximately equal numbers of electrons and holes in these symmetric $\pi$ bands are consistent with the negative thermopower observed [5].

Looking at Figures (1e) and (1f), it is clear that values of $\Delta_z$ greater than 0.085eV render the thermopower completely negative and hyperbolic [20]. Under this condition, the chiral CNT becomes completely n-type material. It was observed in Figure (1e) that at a temperature above 600 K, thermopower becomes zero. A similar observation was made for armchair CNTs [19]. This was attributed to the mirror symmetry of the coexisting electrons and holes in the overlapping $\pi$ bands. An observation made from Figure (1f) shows that when $\Delta_z$ is greater than 0.25eV, increasing $\Delta_s$ does not affect the thermopower.

The thermopower dependence on temperature in the presence and also absence of Laser is sketched and presented as Figure (2). We noted that when the Laser source $E_s$ is between 0 and 4.6 x $10^7$V/m, the thermopower values does not change and this is revealed in Figure 2a where $\alpha_z$ (Laser off) and $\alpha_z$ (Laser on) overlaps. Interestingly, Figure 2b showed a decrease in thermopower when $E_s$ values go beyond 4.6 x $10^7$V/m.

A sketch of thermopower against chiral angle in Figure (3a) showed a rapid decrease in thermopower to a constant value at $7^o$. It therefore indicates that any effect resulting from the thermopower will be extremely small in dependence on the chiral angles beyond $7^o$.

We also sketched thermopower with varying $E_s$ field at a fixed temperature of 300K in Figure (3b). It is interesting to note that as $E_s$ increases, the thermopower shows distinctive peaks. The dependence of thermopower on $E_s$ was found to be oscillatory. Furthermore, thermopower was found to decrease as $E_o$ increases.

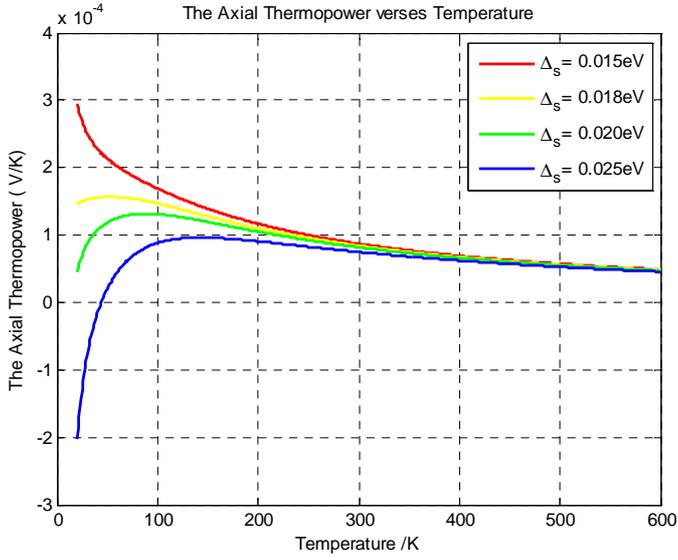

**Figure 1a**: The dependence of $\alpha_{zz}$ on temperature T for $\Delta_s$ equal to 0.015eV, 0.018eV, 0.020eV, 0.025eV, $\Delta_z = 0.015$eV, $E_1 = 5 \times 10^7$V/m, $E_o = .1.38 \times 10^8$V/m.

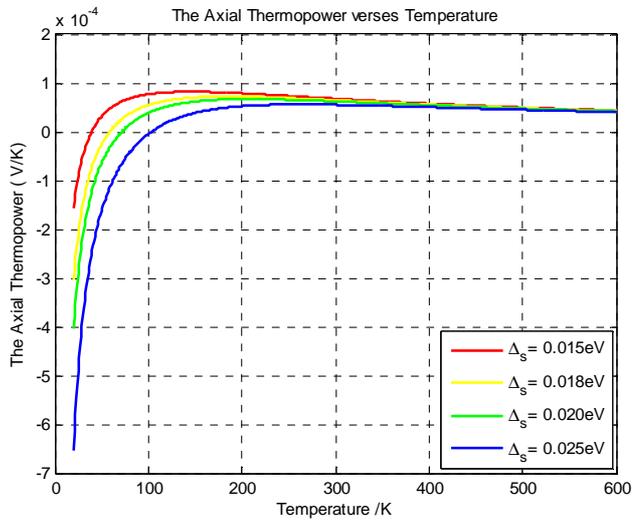

**Figure 1b**: The dependence of $\alpha_{zz}$ on temperature T for $\Delta_s$ equal to 0.015eV, 0.018eV, 0.020eV, 0.025eV, $\Delta_z = 0.024$eV, $E_1 = 5 \times 10^7$V/m, $E_o = .1.38 \times 10^8$V/m..

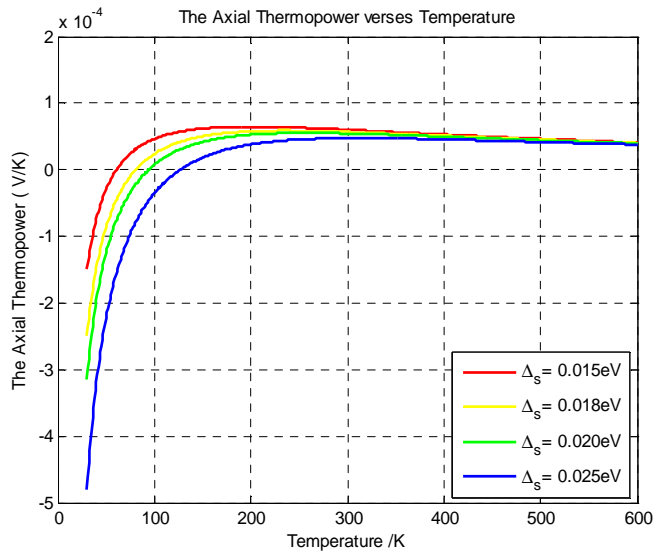

**Figure 1c**: The dependence of $\alpha_{zz}$ on temperature T for $\Delta_s$ equal to 0.015eV, 0.018eV, 0.020eV, 0.025eV, $\Delta_z$ = 0.027eV, $E_1$ =5 x $10^7$V/m, $E_o$ =.1.38 x $10^8$V/m..

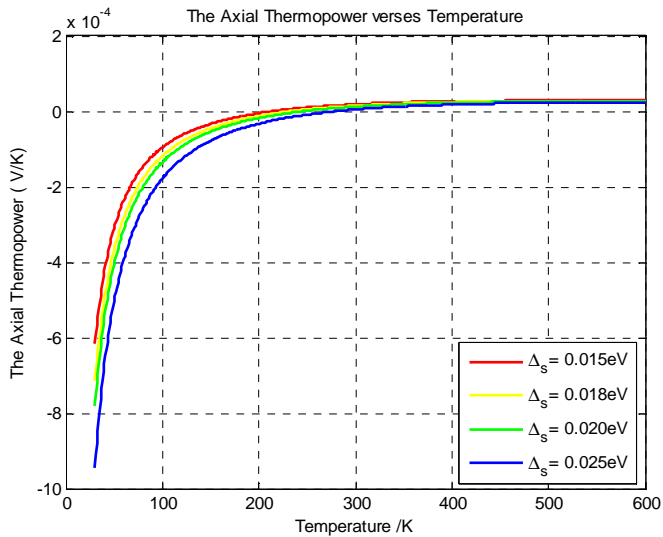

**Figure 1d**: The dependence of $\alpha_{zz}$ on temperature T for $\Delta_s$ equal to 0.015eV, 0.018eV, 0.020eV, 0.025eV, $\Delta_z$ = 0.041eV, $E_1$ =5 x $10^7$V/m, $E_o$ =.1.38 x $10^8$V/m.

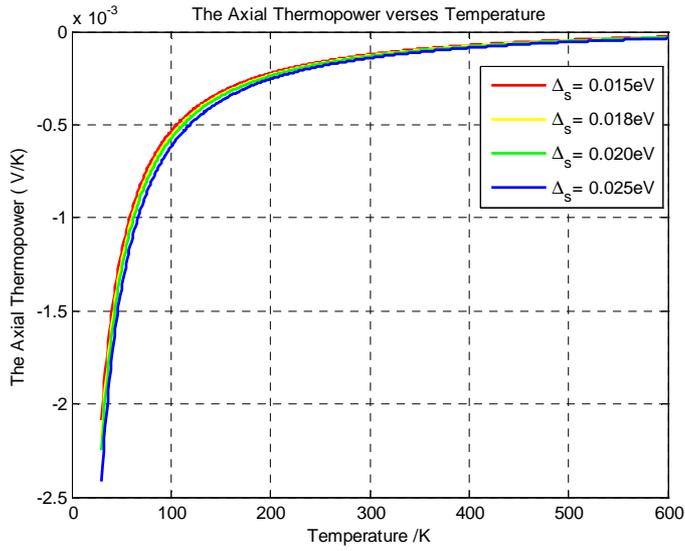

**Figure 1e:** The dependence of $\alpha_{zz}$ on temperature T for $\Delta_s$ equal to 0.015eV, 0.018eV, 0.020eV, 0.025eV, $\Delta_z = 0.085$eV, $E_1 = 5 \times 10^7$V/m, $E_o = .1.38 \times 10^8$V/m.

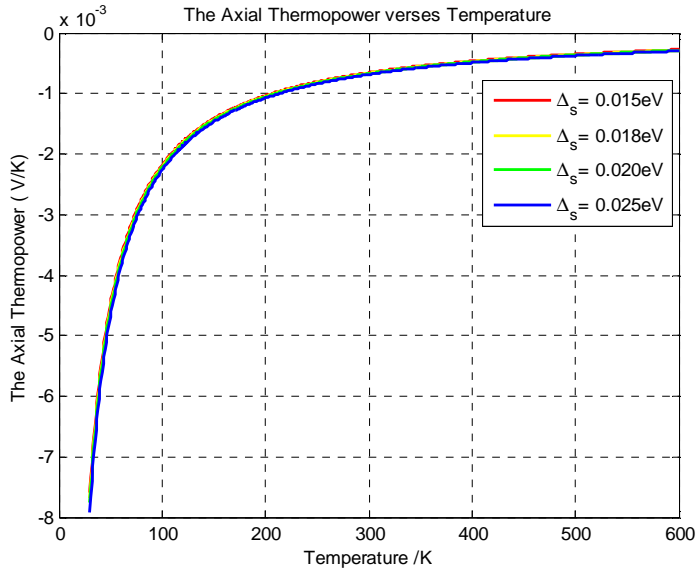

**Figure 1f:** The dependence of $\alpha_{zz}$ on temperature T for $\Delta_s$ equal to 0.015eV, 0.018eV, 0.020eV, 0.025eV, $\Delta_z = 0.25$eV, $E_1 = 5 \times 10^7$V/m, $E_o = .1.38 \times 10^8$V/m..

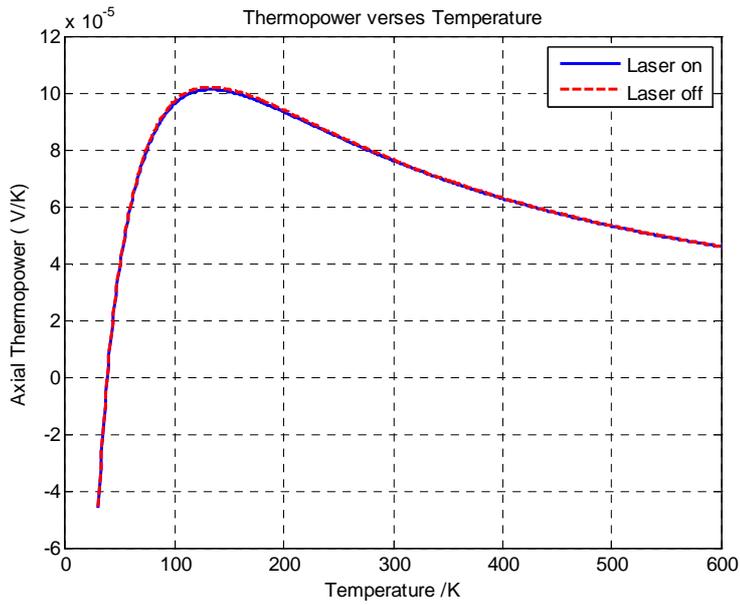

**Figure 2a:** The dependence of $\alpha_z$ on temperature T for $\Delta_s = 0.018\text{eV}, \Delta_z = 0.024\text{eV}$, $E_o = 1.38 \times 10^8 \text{V/m}$. $E_1 = 4.5 \times 10^7 \text{V/m}$ and GCA $\theta_h = 4.0^o$

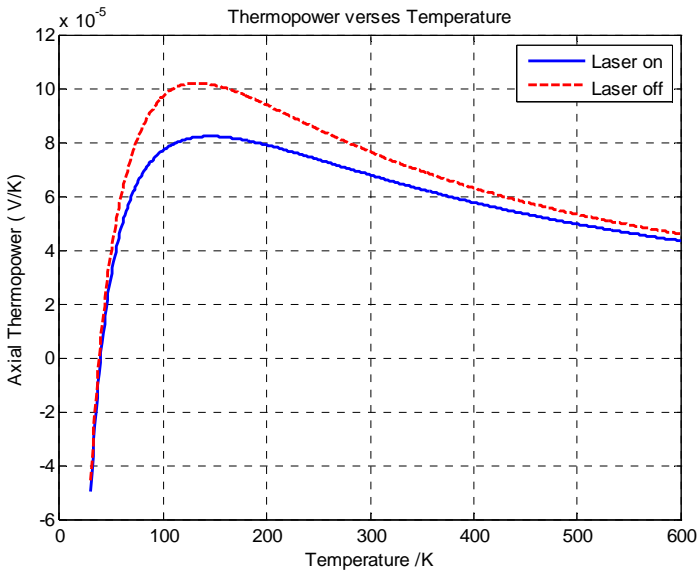

**Figure 2b**

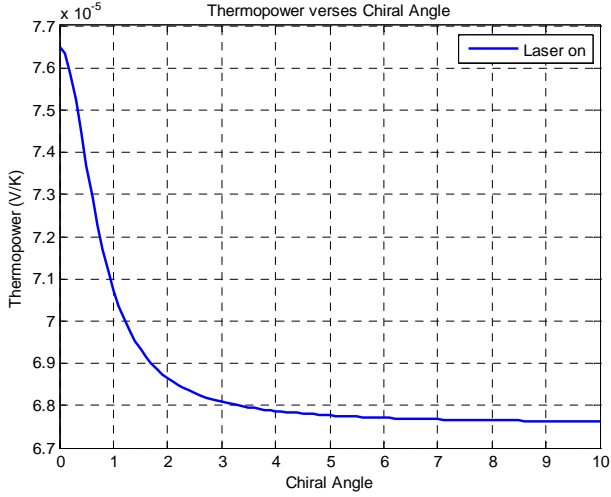

**Figure 3a:** The dependence of $\alpha_{zz}$ on chiral angle at temperature T = 300K, $\Delta_s$ = 0.015eV, $\Delta_z$ = 0.024eV, $E_o$=1.38 X $10^8$ V/m and $E_1$ = 5 x $10^7$V/m.

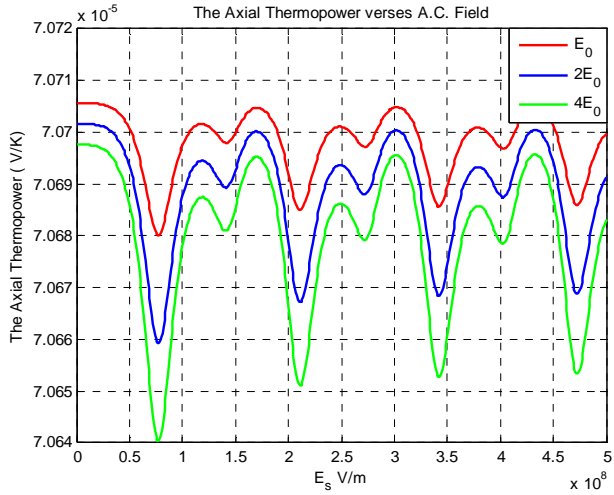

**Figure 3b:** The dependence of $\alpha_{zz}$ on $E_s$ for temperature T = 300K for E = $E_o$, $2E_o$ and $4E_o$, where $E_o$ = 6.9063x $10^7$V/m $\Delta_s$ = 0.018eV, $\Delta_z$ = 0.024eV,

**Conclusions**

The thermopower α of chiral CNT induced with monochromatic laser have been investigated. The chiral CNT parameters $\Delta_s$ $\Delta_z$, $\theta_h$, the d.c. electric field $E_o$ and the laser source $E_s$ were found to have influence on the thermopower α of chiral CNT.

Our results show that the chiral CNT can exhibit semiconducting properties. It became clear that as $\Delta_z$ values increase beyond 0.040eV, the chiral CNT shifts from a p-type to an n-type semiconducting material. We noted that when the Laser source $E_s$ is above 4.6 x $10^7$V/m, the thermopower values decrease. Interestingly, varying the Laser source caused the thermopower of the chiral CNT to oscillate.